\documentclass[12pt]{article}
\usepackage{amssymb}
\usepackage{doublespace}

\def\N{\mathbb N}
\def\a{\alpha}
\def\b{\beta}
\def\d{\delta}
\def\t{\theta}
\def\e{\varepsilon}
\def\g{\gamma}
\def\l{\lambda}

\def\s{\sigma}
\def\L{{\cal L}}

\def\Sumkj{\sum_{j,k=-N}^N}

\def\Prodknoti{\prod_{\tiny \begin{array}{c}k=-N,\\k\not= i\end{array}}^N}
\def\ProdknotI{\prod_{i\in I, \;k\not\in I}}
\def\H{{\cal H}}
\def\be{\begin{equation}}
\def\ee{\end{equation}}
\def\bea{\begin{eqnarray}}
\def\eea{\end{eqnarray}}

\def\ds{\displaystyle}
\def\Bigdot{\begin{picture}(5,5)(0,-2)\circle*{5}\end{picture}}
\def\OSP{$C^{(2)}_{N+1}\,$} 
\def\entete{
\begin{flushright}
{PAR-LPTHE 02-02} \\
{LPTM-cergy}\\
{Janvier 2002}
\end{flushright}
\vspace*{5mm}
}

\def\addJean{LPTHE Paris VI (CNRS-UA 280), Box 126, Univ. Paris 6, 4 Place
Jussieu, F-75252, Paris, France}

\def\monAdd{LPTM Universit\'e de Cergy-Pontoise (CNRS ESA 8089), Neuville III, 5 mail Gay Lussac, Neuville sur Oise, F-95031 Cergy-Pontoise Cedex, France}

\newcommand{\titre}[5]{
\begin{center}
{\bf #1}\\
[10mm]
by\\
[5mm]

#2\footnote{#3} 
and #4\footnote{#5} 
\vskip 1.0in

{\bf ABSTRACT}
\end{center}
\vspace*{3mm}
}

\newtheorem{theorem}{Theorem}
\newtheorem{prop}{Proposition}

\begin{document}
\pagestyle{empty}
\entete

\titre{ \OSP Ruijsenaars-Schneider 
models}{J. AVAN}{\addJean}{G. ROLLET}{\monAdd}

\begin{spacing}{1.2}
\noindent 
We define the notion of \OSP Ruijsenaars-Schneider models and
construct their Lax formulation.
They are obtained by a particular folding of the $A_{2N+1}$ systems.
Their commuting Hamiltonians are linear combinations of  
Koornwinder-van Diejen ``external fields'' Ruijsenaars-Schneider 
models, for specific values of the exponential one-body couplings
but with the most general $2$ double-poles structure as opposed
to the formerly studied $BC_N$ case. Extensions to the elliptic potentials
are briefly discussed.

\vfill
\end{spacing}
\newpage
\setcounter{page}{1}
\pagestyle{plain}
\renewcommand{\baselinestretch}{1.5}
\begin{spacing}{1.5}

\section{Introduction}

We wish to present here an explicit Lax formulation for a subclass of one-body extensions of the
classical integrable Ruijsenaars-Schneider (RS) models~\cite{RS} quantized in ~\cite{RS3}. The
quantum version of these extensions was formulated initially by van Diejen~\cite{VD1,VD2,VD3} 
starting from the pioneering works of Koornwinder on algebraic formulations of extensions of 
MacDonald polynomials~\cite{Ko}. This formulation, and the proof of quantum integrability
relied upon analytical arguments using the newly constructed Koornwinder polynomials as a basis
for the wave function Hilbert space~\cite{VD2}. A quantum formulation for the elliptic 
hamiltonians, conjectured in~\cite{VD3}, was given in a series of works by Komori et 
al.~\cite{Kom} using a corner-transfer matrix method combining functional representations
of both bulk quantum $R$-matrix and boundary reflection $K$-matrices. General one-body 
extensions 
of the difference RS operators were thus built, limits of which could be identified with the 
one-body extensions of differential Calogero-Moser (CM) Hamiltonians constructed by 
Inozemtsev~\cite{Ino} and associated to $BC_N$ lattices~\cite{BSa}.
\vskip0.10in
It must be noticed at this point that no clear classical limit of this construction exists,
entailing as it does a coordinate-permutation operator in the functional representation of
the quantum bulk $R$-matrix and a coordinate-reflection operator in the quantum reflection
$K$-matrix. In fact, no full classical Lax formulation exists for the most
general Koornwinder-van Diejen (KvD) Hamiltonians; a first step in this direction was our
identification of the $BC_N$ Ruijsenaars-Schneider Hamiltonians as linear combinations of 
particular KvD hamiltonians~\cite{AR2}. In addition the classical $r$-matrix was obtained
in this case, providing an interesting example of a dynamical dependance in {\it both}
sets of dynamical variables, rapidities and positions.
\vskip0.10in
Our purpose is to describe a classical Lax formulation of a more general subclass of KvD
potentials; more precisely the one-body part of the potential will exhibit the same 2
double-pole
dependence as the full KvD; the coupling constants however, identified with the residues,
still depend on one single coupling. This however marks a qualitative advance compared to the
previous construction of $BC_N$ RS model, since one there had only one double and 
{\it one single} pole. 
\vskip0.10in
As in the previous case we will rely on a consistent $Z_2$ folding procedure applied to a 
suitable $A_n$ RS model to obtain new classical integrable systems and their Lax formulation.
The first suggestion of constructing (in that case) $BC_N$ and $C_N$ RS models by 
$Z_2$ folding of $A_{2N (+1)}$ RS model came from Ruijsenaars himself~\cite{RS2} and was later explicited 
at the level of the Lax formulation~\cite{CH,CHY2}. A related programme was applied to the
simpler case of CM models~\cite{BSa}, using folding procedures on the phase-space
variables and twistings of the underlying Lie algebras to obtain general one-body extensions
(see also~\cite{Ino}).
 
We here establish that this more general class of KvD potentials~\cite{VD1}
(restricted to the hyperbolic
case for the sake of simplicity) is obtained as linear 
combinations of pure RS Hamiltonians which may be described in terms of the root lattice
of the twisted affine superalgebra
$OSp(2|2N)^{tw}$ or \OSP. The lattice characterizing the form of the potential indeed
exhibits the shift of shortest roots by $\frac{1}{2}$ the derivation, characteristic of the \OSP
root lattice~\cite{FF}
\footnote{We are indebted to Luc Frappat for providing this reference and the identification of \OSP}.
At this time however we lack a deeper interpretation of the occurrence of this particular
lattice.

The problem of finding the Lax
formulation for the most general four-coupling hyperbolic KvD potentials remains open
at this point, but we conjecture that this subclass, exhibiting as it does the full pole 
structure, constitutes the best starting point for eventual achievement of this construction. 

The detailed analysis of elliptic generalizations will be left for later studies. 
However it is established here that the most general $8$ pole structure found in~\cite{Kom}
(degenerating to a $4$ double-pole structure in the classical limit)
may be similarly obtained by a generalized folding operation  and the previous conjecture on
the hyperbolic case is therefore extendable to the elliptic case. 

A final comment regarding the denomination ``\OSP RS model'' used here. The qualitative difference 
between
the class of potentials obtained from the initial $BC_N$ folding and this \OSP folding,
i.e. the occurrence of a $2$ double pole structure in the one-body term instead of a double-pole
times single-pole structure, vanishes in the CM limit where poles {\it add} instead of
{\it multiply}. It follows that the distinction between $BC_N$ and \OSP CM models is
essentially non-existent at least at the level of Hamiltonians, and 
the litterature (see for instance~\cite{BSa}) rightly 
characterizes only ``$BC_N$ type''  models. 
\section{The \OSP Ruijsenaars-Schneider model}
\subsection{Invariant subspaces of the $A_n$ RS dynamics}

Presentation of the \OSP RS model is easily done in a more general framework where one 
constructs subsets of the phase space for an $A_n$ RS hierarchy so that they be invariant
under a subclass of the flows generated by some specific combinations of the Hamiltonians.
We now recall the definition of the $A_n$ RS hierarchy, for any $n$.

The canonical variables are a set of rapidities $\{ \t_i, i=1\cdots n+1\}$
and conjugate positions $q_i$ such that $\{ \t_i, q_j\} = \delta_{ij}$.
The Hamiltonians are initially defined as:
 
$ h_l  = \ds \!\!\!\!\!\!\!\sum_{I \subset \{ 1 \cdots n+1 \},|I| = l} \!\!\! e^{- \beta \t_I} f_I 
\quad \mbox {where}\quad f_I = \!\!\!\!\ProdknotI f(q_i - q_k)^{1/2} \nonumber \quad \mbox{and}\quad \t_I = \sum_{i \in I} \t_i $.

Function $f$ may take different forms (rational, hyperbolic, trigonometric), resp:
$$ f(q)  =  1 - {g^2\over q^2} \,\, ; \quad
f(q)  =  1 - {{\rm sinh}^2 \g \over {\rm sinh}^2 {\nu q\over2}}\,\, ;\quad
f(q)  =  1 - {{\rm sin}^2 \g \over {\rm sin}^2 {\nu q\over2}}\,\, $$
The most general elliptic case where $f (q)  =  \left( \lambda + \nu {\cal P} (q) \right) ,
\,\,{\cal P}$ being Weierstrass function
will not be fully explicited here. Our construction may however be extended to 
it with due modifications to be discussed in the appropriate place (see Remark 2 in this section and Comment 2 in
Section 5).

The trigonometric and hyperbolic cases define the same model at least
locally up to a redefinition of the parameters 
(the global structure of trigonometric {\it vs.} hyperbolic models is however quite different,
owing to qualitatively distinct topological properties, as can be seen for instance in~\cite{Go}). 
The rational case is obtained by an easy limit procedure
from one of these models. We shall therefore consider in the following only 
the hyperbolic model.

Let us note that $f(q)=v(q)\,v(-q)$, with :
$v(q)={{\rm sinh}({{\nu q}\over{2}} +\g) \over {\rm sinh}{\nu q\over2}}
=\l^{-1/2}{{z-\l}\over{z-1}}$.
These functions are thus  rational fonctions of an exponential variable,
$z=e^{\nu\,q}$, defining $\l=e^{-2\,\g}$.
The square root is defined here with a cut on the real negative axis.

Both functions $f$ and $v$ are periodic with period $T \equiv \frac{2i \pi}{\nu}$.
One introduces now a better adapted (although possibly degenerated) set of Hamiltonians as:
\be\label{ham2}
K_0 = \frac{h_{n+1} + h^{-1}_{n+1}}{2} \; ; \; \mbox{for} \;l= 1 \cdots \left[ \frac{n+1}{2} \right], \; K_l   =  h_l, \;
K_{-l}   = {\frac {h_{n+1-l}}{h_{n+1}}}
=\sum_{|I| = l}  e^{\beta \t_I} f_I 
\ee

The negative-index Hamiltonians are in fact obtained from traces of {\it negative powers} of the 
RS Lax matrix, using its remarkable Cauchy structure~\cite{RS} to connect $L^{-1}$ with $L^{t}$.
One then considers any idempotent bijection $\sigma$ over the index set $\{1 \cdots n+1\}$, separating them
into invariant singlets $\sigma(i) = i$ and doublets $\sigma(i) = j \, , \, \sigma(j) = i \, , \, j \neq i$.
It is easily shown that:
\begin{prop}
For any idempotent bijection $\sigma$, the manifolds defined by:
\vskip -.5cm
\be\label{fold}
\forall i, q_i + q_{\sigma(i)} = 0 (T) \; ; \; 
 \t_i + \t_{\sigma(i)} =0 (\frac{2 i \pi}{\beta})
\ee
\vskip -.25cm
are kept invariant by the evolution generated by the Hamiltonians ${{K_l + K_{-l}}\over{2}} \equiv \H_l$.
\end{prop}

\noindent{\bf Proof:}
The evolution equations are:
$\ds \{ q_j, \H_l \} = \b\,\sum_{I \ni j}{{e^{\b \t_I} - e^{- \b \t_I}}\over{2}} f_I\; \mbox{and} \; \{ \t_j, \H_l \} =$
\be\label{evol}
\sum_{I \ni j}{{e^{\b \t_I} + e^{-\b \t_I}}\over{2}} f_I \sum_{k \not\in I} (ln \, f^{1/2})'(q_j - q_k)
-\sum_{I \not\ni j}{{e^{\b \t_I} + e^{-\b \t_I}}\over{2}} f_I \sum_{i \in I} (ln \, f^{1/2})'(q_i - q_j).
\ee

Invariance of the manifolds (\ref{fold}) under (\ref{evol}) is 
straightforwardly obtained from the parity and $T$-periodicity 
of the two-body potential function $f$; the explicit 
$\t$ parity and $\frac{2 i \pi}{\beta}$-periodicity of $\H_l$; and the bijectivity
of $\sigma$ allowing adequate index redefinitions.

The number of $\s$-stable indices is a priori arbitrary. However if there are more than two such
indices, one will unavoidably have exact equality mod. $T$ of at least two position variables $q$ and
the invariant submanifold will actually lie in the singularity hyperplanes $q_i = q_j (T)$. This case
must therefore be eliminated from our discussion, which leaves us with only three possibilities.

{\bf Case 1:}
For $n$ even (odd number of sites), one may only consider the case of one stable index. This 
construction gives us the $BC_{\frac{n}{2}}$ RS model~\cite{RS2,CH,AR2}.

{\bf Case 2:}
For $n$ odd (even number of sites), one may first consider the case of zero stable index.
This leads to the $C_{\frac{n+1}{2}}$ case~\cite{RS2,CHY2}.

{\bf Case 3:}
For $n$ odd (even number of sites), one may then consider the case of two stable indices
hereafter denoted $0$ and ${\bar 0}$.
The only non-singular choice for values of the two fixed position coordinates is then $q_0 = 0$ and $q_{\bar{0}}
= T/2$ up to trivial permutation. Connection to the superalgebra \OSP 
with $n \equiv 2N+1$ will be established 
presently.

{\bf Remark 1:}
Regarding the rapidities, we will here restrict ourselves for the sake of simplicity
to the choice $\t_0 = \t_{\bar 0} = 0 
(\frac{2 i \pi}{\beta})$. This implies immediately that the restriction of $\H_0$ to (\ref{fold})
is identically $1$. We shall comment at the end (Comment 3, Section 5) on the implications
of other possible choices.

We now establish the important properties of this consistent restriction.
We first introduce
a better adapted set of conjugate variables (for the full phase space) defined to be
$\{\frac{1}{{\sqrt 2}} (q_i - q_{\sigma(i)}), \frac{1}{{\sqrt 2}} (\t_i - \t_{\sigma(i)}) 
\}$; $\{\frac{1}{{\sqrt 2}} (q_i + q_{\sigma(i)})
 \equiv q_i^{\perp}, 
\frac{1}{{\sqrt 2}} (\t_i + \t_{\sigma(i)}) \equiv \t_i^{\perp}) \}\, , \; 
 \forall i \, , \,  \sigma(i) \neq i$; finally $q_0 \equiv q_0^{\perp},  \t_0 \equiv \t_0^{\perp},
q_{\bar 0}- \frac{T}{2} \equiv q_{\bar 0}^{\perp},  \t_{\bar 0} \equiv \t_{\bar 0}^{\perp}$.

\noindent {\bf Corollary 1a:}
{\it
These submanifolds are obviously endowed with a symplectic form $\{ \, \}_{\rm rest}$
where conjugate variables are respectively the restrictions of $\frac{1}{\sqrt 2}(q_i - q_{\sigma(i)})$ and 
$\frac{1}{{\sqrt 2}} (\t_i - \t_{\sigma(i)})$ for $i$ such that $\sigma(i) \neq i$.}

\noindent{\bf Corollary 1b:}
{\it
The Hamiltonians $\H_l$, restricted to the submanifolds (\ref{fold}) endowed with the
symplectic form $\{ \, \}_{\rm rest}$ build an integrable hierarchy.}

\noindent {\bf Proof:}
An obvious rewriting of Proposition $1$ states that the Hamiltonians $\H_l$
expanded around the values $\t_i^{\perp} =0, q_i^{\perp} =0$, have no linear term in 
$\t_i^{\perp} , q_i^{\perp} $: indeed such terms would trigger a non-trivial
dynamics of $\t_i^{\perp} , q_i^{\perp} $ on (\ref{fold}). 
This leads us to define $\ds \H_l^{rest} \equiv 
\H_l (\{\frac{1}{{\sqrt 2}} (q_i - q_{\sigma(i)}), \frac{1}{{\sqrt 2}} (\t_i - \t_{\sigma(i)}) ,
q_i^{\perp}=0, \t_i^{\perp}=0)$
and now $\H_l = \H_l^{rest} +$ quadratic terms in $q_i^{\perp},\t_i^{\perp}$

In addition the $\H_l$ Poisson-commute by construction. This
is now expressed as :
\vskip -.5cm
\[
\{ \H_a, \H_b \}_{{\rm full}} = 0 \equiv \{ \H_a^{rest}, 
 H_b^{rest} \}_{{\rm full}} + \t_i^{\perp} \times \cdots
+ q_i^{\perp} \times \cdots
\]
\vskip -.25cm
which implies in particular:
  $ \{ \H_a^{rest}, 
 H_b^{rest} \}_{{\rm full}}= 0. $

Finally one has $ \{ \H_a^{rest}, 
 H_b^{rest} \}_{{\rm full}}\equiv \{ \H_a^{rest}, 
 H_b^{rest} \}_{{\rm rest}} = 0
$ by definition of the 
restricted  symplectic structure on the submanifolds, thereby ending the proof.

For the sake of simplicity we shall from now on drop the ``rest'' index in $\H_l$.

{\bf Remark 2:} 
In the elliptic case the potential function is biperiodic, hence one may allow 
for $4$ fixed coordinates since two independent periods are available to define the invariant
subspaces (replacing the single period congruence parameter $T$ in (\ref{fold})). 

We now formulate:
\begin{prop}
The Hamiltonians $K_l$ and $K_{-l}$ are identical, and therefore 
equal to $\H_l$, on the submanifolds (\ref{fold}).
\end{prop}

The proof is a trivial consequence of the bijectivity of $\sigma$; the fact that 
on the reduced submanifolds (\ref{fold}) one has up to a full period $\t_{\sigma(i)} = -\t_i$
and $q_{\sigma(i)} = -q_i$ for {\it all} indices; and the parity and
periodicity properties of the potential $f$. 
\vskip0.10in
The immediate crucial consequence is that the restriction of the Hamiltonians $\H_l$
 can be rewritten as:
$\H_l = P_l ({\rm Tr} L^m)$
where $P_l$ is the $l$-th Newton polynomial (sum of rank $l$ minors) and 
$L$ is the restriction of the Lax matrix of $A_{2N+1}$ RS model to the submanifolds
(\ref{fold}).
We are therefore provided with a Lax representation for the integrable hierarchy $\H_l$.
\vskip0.10in
We now discuss the particular dependance of the potential terms in the Hamiltonians, in order to 
justify the claimed connection to the twisted affine superalgebra \OSP.
Specific connection is the following: the fixing of $q_{\bar 0}$
to the value $T/2$ introduces an imaginary shift in the lattice describing the
position dependance of the potential function, turning it into the 
root lattice of \OSP .

More precisely the original dependance of the potential
function was in the variables $q_i - q_j, i,j \in \{ 1 \cdots 2N+2 \}$,
associating it with the root lattice of $A_{2N+1}$.  After this particular folding we now
get a dependence in $q_i - q_j; q_i + q_j; q_i; 2q_i; q_i + T/2; i,j 
\in \{ 1 \cdots N \}$. This immediately leads to seek for a lattice
with three root lengths together with the simultaneous existence of shifted
and unshifted shortest roots. By exploration~\cite{FF} the only possibility is \OSP.
 
These variables may now
be rexpressed as scalar products $<\alpha.Q>$
where $Q$ is the $(N+1)$-component vector in the dual of the Cartan algebra
with coordinates $q_1 \cdots q_N$ on the dual of the Cartan algebra of the simple Lie algebra
and $T$ along the direction of the derivation generator $d$ of the full
Cartan algebra; $\alpha$ is any
root of \OSP \cite{FF}, exhibiting in particular the shift of
the shortest roots by $\frac{1}{2}$ the derivation generator characteristic of \OSP. It remains
at this time still a formal connection, and we have no interpretation
of the occurence of a twisted affine superalgebra in this context.

The pure $BC_N$ folding (case 1) differs qualitatively in that the one-body part of
the potential for the first Hamiltonian (i.e. linearly dependent upon single exponentials of 
rapidities) does not contain a double pole at half period due to the absence
of the extra dependance in $ q_i + T/2$. In the non-relativistic CM
limit however, the one-body potential in the first (quadratic) Hamiltonian
obtained by {\it both} foldings exhibits double poles at half-period (occurring from the
folding of terms sinh$(q_i - q_j )^{-2}$ at $q_j = -q_i$ giving in particular
a term cosh$(q_i)^{-2}$ by doubling of the sinh; or at $q_j = \frac{T}{2}$); and double poles 
at integer period (occurring from the folding of terms sinh$(q_i - q_j )^{-2}$ at $q_j = 0$).
The technical reason is the multiplicative nature of the potential terms in a RS model as 
opposed to the additive nature thereof in a CM model, and is thus related
to the fundamental difference between RS models realized~\cite{GN} on a Heisenberg 
double~\cite{STS,Dr,Al} and CM models realized on a cotangent bundle~\cite{OP}.
Actually, the only difference between $BC_N$ and \OSP foldings of CM models
lies in the value of the residues (coupling constants) which
in any case may be extended to take any value, hence it turns out to be irrelevant.

\section{The \OSP Lax operator and $r$-matrix}

We shall from now on use the notation $-i$ instead of $\sigma(i)$, defining that $- {\bar 0} = {\bar 0}$.
The Lax formulation of \OSP RS
system is therefore obtained as a folding of the A$_{2N+1}$ case~\cite{RS2}
explicitely built as follow:

We first label the $2\,N+2$ rapidities $\{ \t_i, i=-N\cdots 0, {\bar 0}, \cdots N\}$ and conjugate positions  $\{ q_i, i=-N\cdots  0, {\bar 0}, \cdots N\}$. 
Independent phase space variables on the restricted manifolds are here 
chosen to be $q_i, \t_i, i \in \{ 1 \cdots N \}$. 
Note that on the restricted manifold the Poisson structure $\{ \}_{rest}$
in terms of these variables reads $\{ \t_i, q_j\} = \frac{1}{2}\delta_{i,j}$
which will be responsible for an overall $\frac{1}{2}$ factor in the Poisson structure.
This factor will be omitted from now on, corresponding to a normalization of the Poisson bracket as
$\{ \t_i, q_j\} = \delta_{i,j}$.
One thus identifies $\t_i=\e_i\,\t_{|i|}$ and $q_i=\e_i\,q_{|i|} + \delta_{i, {\bar 0}} \frac{T}{2}$ with
$\forall i\in\{ 1 \cdots N \},\,\e_i=1=-\e_{-i}$ and $\e_0=\e_{\bar 0}=0$.

The Lax matrix for the A$_{2N+1}$ cases reads:
$ \ds \L = \sum_{i,j=-N}^N \, \L_{ij} \, e_{ij}$ 
where $\L_{ij}(q_{-N},...,q_{N},\t_j)=c(q_i-q_j)\,e^{-\b\,\t_j}\,f_{\{j\}}$,
$\{ e_{ij} \}$ is the standard basis for $(2 N+2) \times (2 N+2)$ matrices and
$c(q)={{\rm sinh}\g \over {\rm sinh}({{\nu q}\over{2}} +\g)}
=(1-\l)\,{{z^{1/2}}\over{z-\l}}$.
\vskip .25cm
The Lax matrix for the \OSP Ruijsenaars-Schneider systems then reads:
\vskip -1cm
\be
L =\sum_{i,j=-N}^N \, L_{ij} \, e_{ij} \quad \mbox{with} \quad
L_{ij}=\L_{ij}(-q_N,...,-q_1,0, \frac{T}{2}, q_1,...,q_N,\e_j\,\t_{|j|})\label{LaxOSP}
\ee
\vskip -.25cm
It is now well-known that the Lax operator $\L$ satisfies the
quadratic fundamental Poisson bracket~\cite{Su}:
$\left\{ \L \stackrel{\otimes}{,}\L\right\}=
\L\otimes\L\;a_{1}-a_{2}\;\L\otimes\L+\L_{2}\,s_{1}\,\L_{1}-\L_{1}\,s_{2}\,\L_{2}$,
where \ $\L_{1}=\L\otimes \emph{1}$, $\L_{2}=\emph{1}\otimes \L $ and the 
quadratic structure coefficients read:
 $ a_{1} =a+w, \, s_{1}=s-w, \,  
a_{2} = a+s-s^{\pi }-w,\,s_{2}=s^{\pi }+w.$
As usual, for any matrix $\ds M=\!\!\!\sum_{ijkl=-N}^N \!\!\!\! M_{ijkl} \, e_{ ij} \otimes e_{kl}$ the operation $\pi$ is defined by:
$M^{\pi}_{ijkl}= M_{klij}$.
Matrices $a,s,w$ take the form~:
\vskip -.5cm
\[ a =-\a\!\!\!\!\Sumkj \!\!\!a_{jk}\,e_{jk}\otimes e_{kj},\;
s =\a\!\!\!\!\Sumkj \!\!\!s_{jk}\, e_{jk}\otimes e_{kk},\;
w =\a\!\!\!\!\Sumkj \!\!\!a_{jk}\, e_{jj}\otimes e_{kk}\;
\mbox{with}\; \a=\b\,{{\nu}\over{2}}\]
and $\ds a_{jk}=(1-\d_{j,k})\,\mathrm{coth}{{\nu}\over{2}}(q_{j}-q_{k})
=(1-\d_{j,k})\,{{z_j+z_k}\over{z_j-z_k}},\;
s_{jk}=\frac{(1-\d_{j,k})}{\mathrm{sinh}{{\nu}\over{2}}(q_{j}-q_{k})}$.

Remember that the most general structure of Poisson bracket
for a Lax operator of a Liouville-integrable system is a linear one~\cite{BV}:
$\left\{ L \stackrel{\otimes}{,}L\right\}=[r,L_1]-[r^{\pi},L_2]$.
The quadratic form corresponds to the general case~\cite{LP} where the 
$r$-matrix itself assumes a linear dependence in $L$ of form: 
$r=b\,L_2+L_2\,c$ with $b$ and $c$ arbitrary matrices yielding~:
$ a_1=c^{\pi}-c,\;a_2=b^{\pi}-b,\;s_1=c+b^{\pi} \; \mbox{and} \; s_2=s_1^{\pi} $.

We now straightforwardly extend the previous
computation~\cite{AR2} of the $BC_N$ Ruijsenaars-Schneider $r$-matrix. The Lax  
operator also satisfies a quadratic fundamental Poisson 
bracket, again exhibiting the dependence of the structure matrices $a$ and $s$
on both sets of dynamical variables. The occurrence of two zero-type indices 
$0$ and ${\bar 0}$ requires extra caution at one particular place, eventually leading to 
supplementary signs in the formula for factors of the $r$-matrix. Note finally that this 
derivation is equivalently applicable to the $C_N$-type RS Lax matrix, this time by 
altogether removing the zero-indices.

The $r$-matrix structure is again completely defined by a quadratic Poisson 
bracket with  $a_1,\,a_2,\,s_1$ and $s_2$ changed into:
$a_1\,\rightarrow \tilde{a}_1=a_1,\;
a_2\,\rightarrow  \tilde{a}_2=\tilde{a}_1+\tilde{s}_1-\tilde{s}_2
=a_2+\tau^{\pi}-\tau,\;
s_1\,\rightarrow  \tilde{s}_1=s_1+\sigma+\tau^{\pi},\;
s_2\,\rightarrow  \tilde{s}_2=\tilde{s}_1^{\pi}
=s_2+\sigma^{\pi}+\tau=s_2+\sigma+\tau$, with:
\[
\tau =
\a\!\!\!\!\sum_{i,k,l=-N }^N\!\!\!\!L_{k-i}\,L_{-il}^{-1}\,({{u_{-i}}\over{2}}-t_{k-i})
\,e_{ii}\otimes e_{kl},\;
\s = \a\!\!\!\sum_{i,j=-N }^N\!\!\!
\left (\d_{i,j}\,s_i-(1-\d_{i,j})\,a_{ij}\right )
{{{\cal A}_j}\over{{\cal A}_i}}\,
e_{ij}\otimes e_{-j-i},
\]
defining 
$\ds t_{ij}=\mathrm{coth}({{\nu}\over{2}}(q_{i}-q_{j})+\g)
={{z_i+\l\,z_j}\over{z_i-\l\,z_j}}$,
$\ds s_i={{1+\l}\over{1-\l}}\,+\,\sum_{m=-N }^N\,{{t_{mi}+t_{im}}\over{2}}$,
$\ds u_i=\sum_{k=-N}^{N} \!\!\!2 a_{ik}+t_{ki}-t_{ik}$ and
$\ds {\cal A}_i =\!\!\!\!\Prodknoti\!\!\!({{z_i-\l\,z_k}\over{\l\,z_i-z_k}})^{1/2}\,e^{-\b\,\t_i} e^{i \pi \delta_{i, {\bar 0}}}$.

Occurrence of the final sign factor  $e^{i \pi \delta_{i, {\bar 0}}}$ is the only qualitative modification
induced by the supplementary fixed index ${\bar 0}$. In addition the full $r$-matrix gets an overall
$\frac{1}{2}$ factor due to the normalization of the restricted Poisson bracket.

As in the $BC_N$ case this quadratic $r$- matrix structure is
fully dynamical, depending both
on the positions $q_i$'s and rapidities $\t_i$'s.

\section{Koornwinder-van Diejen versus \OSP Ruijsenaars-Schneider Hamiltonians}

As in the $BC_N$ case the Hamiltonians $\H_l$ generated by
traces of powers of the \OSP Lax matrix~(\ref{LaxOSP}) can be rexpressed in a more 
interesting form as:
\bea\label{calHl}
{\cal H}_l&=&\hskip -.5cm
\sum_{\tiny \begin{array}{c}
J\subset \{1..N\},\,|J|\le l\\
\e_j=\pm 1, \, j \in J
\end{array}} {\cal U}_{J^c,l-|J|}\quad e^{-\b \t_{\e J}}
\,f_{\e J}, \; \mbox{with} \quad \e J \equiv \{\e_j |j|, j \in J \}
\quad \mbox{and}\nonumber\\
{\cal U}_{K,p}&=&\hskip -.5cm
\sum_{\tiny \begin{array}{c}
S\subset {\cal A}_{K}=K \bigcup -K \bigcup \{0, {\bar 0} \}\\
S=-S,\,|S|=p
\end{array}}\hskip -.25cm
\prod_{\tiny \begin{array}{c} 
s\in S\\ k\in {\cal A}_{K}\backslash S
\end{array}}\hskip -.25cm f^{1/2}(q_s-q_k)
=\hskip -.5cm
\sum_{\tiny \begin{array}{c}
S\subset {\cal A}_{K}\\
S=-S,\,|S|=p
\end{array}}\hskip -.25cm
\prod_{\tiny \begin{array}{c} 
s\in S\\ k\in {\cal A}_{K}\backslash S
\end{array}}\hskip -.25cm v(q_s-q_k).
\eea

We now recall the form of classical  Koornwinder-van Diejen Hamiltonians~\cite{VD3}:
\be\label{Hl}
H_l=
\sum_{\tiny \begin{array}{c}
J\subset \{1..N\},\,|J|\le l\\
\e_j=\pm 1, \, j \in J
\end{array}} \hskip -.25cm U_{J^c,l-|J|}\quad e^{-\b \t_{\e J}}\quad
V^{1/2}_{\e J; J^c}\quad V^{1/2}_{-\e J; J^c},
\ee
where, after some rearrangements required to reintroduce the indices
$0, {\bar 0}$:
\bea\label{polyn}
V_{\e J; K}&=&\prod_{j \in \e J} w_r(q_j)
\!\!\!\!\!\!\!\prod_{\tiny \begin{array}{c}
j \in \e J \\ k \in {\cal A}_K \bigcup -\e J
\end{array}} \hskip -.5cm v(q_j-q_k),
\; \mbox{with} \quad 
w_r(q_j)={{w(q_j)}\over{v(2\,q_j)\,v(q_j)\,v(q_j - \frac{T}{2})}}\nonumber\\
\mbox{and} \quad U_{K,p}&=&(-1)^p 
\sum_{\tiny \begin{array}{c}
\e I \subset {\cal A}_{K}\\
|I|=p
\end{array}} 
\prod_{i \in \e I} w_r(q_i)
\prod_{\tiny \begin{array}{c}
i, i' \in \e I\\i < i'
\end{array}} \hskip -.25cm
{{v(-q_i-q_{i'})}\over{v(q_i+q_{i'})}}
\prod_{\tiny \begin{array}{c} 
i\in \e I\\ k\in {\cal A}_{K}\backslash \e I
\end{array}} \hskip -.5cm v(q_i-q_k).
\eea
The potentials $w$ are particular functions explicited in~\cite{VD1}
and may be interpreted physically as an interaction with some external field.

Direct computation yields:
$\ds V_{\e J; J^c}\; V_{-\e J; J^c}=
\prod_{j \in \e J} w_r(q_j)\,w_r(-q_j)\,f^2_{\e J}$.

Setting $w_r(q_j)=1$, that is $w(q_j)=v(2\,q_j)\,v(q_j)\,v(q_j -\frac{T}{2})$, which is an admissible choice 
according 
to~\cite{VD3}, $H_l$~(\ref{Hl}) takes actually the same form as 
${\cal H}_l$~(\ref{calHl}), up to
the crucial change of ${\cal U}_{K,p}$ into $U_{K,p}$.
They are generally not equal, except for $p=0$, where trivially:
$U_{K,0}=1={\cal U}_{K,0}$.
When $p=1$, one gets:
\[
U_{K,1}=-\sum_{i \in {\cal A}_{K}\backslash \{0, {\bar 0}\}}
\prod_{\tiny \begin{array}{c}  k\in {\cal A}_{K}\\ k\not=i\end{array}} \hskip -.25cm v(q_i-q_k) \quad \mbox{and} \quad 
{\cal U}_{K,1}=
\prod_{\tiny \begin{array}{c}  k\in {\cal A}_{K}\\ k\not=0\end{array}}\hskip -.25cm v(q_k)
+  \prod_{\tiny \begin{array}{c}  k\in {\cal A}_{K}\\ k\not={\bar 0}\end{array}}\hskip -.25cm v(\frac{T}{2} -q_k).
\]
Evaluation of a suitable contour integral as in~\cite{AR2} gives
the Liouville-type functional identity: $\ds \sum_{i \in {\cal A}_{K}}
\prod_{\tiny \begin{array}{c}  k\in {\cal A}_{K}\\ k\not=i\end{array}} \hskip -.25cm 
v(q_i-q_k)=$${{\sinh \g(2\,|K|+2)}\over{\sinh \g}}$, that is:
$U_{K,1}={\cal U}_{K,1}-{{\sinh \g(2\,|K|+2)}\over{\sinh \g}}$.

We now recall the general theorem established in~\cite{AR2}:

\begin{theorem}
Let $q_i$ and $\t_i$, $i\in \N$, be a set of conjugated variables
such that $\{ \t_i, q_j\} = \delta_{ij}$.
Let $I$ and $K$ be  arbitrary finite sets of indices included in $\N$.
Assume the existence of a set of complex functions  $u_{K,p}$ depending 
upon the set of
indices $K$  and a natural integer $p$, and of another set
of complex functions  $v_{\e J,I}$ depending upon the sets of
indices $J$ and $I$ ($J\subset I$) and a $|J|$-uple of signs  $\e=(\e_j,\,j\in J)$,
such that:

\Bigdot $u_{K,p}$ and $v_{\e J,I}$ be independent of the rapidities $\t_i$s.

\Bigdot $u_{K,0}=1$, $v_{\emptyset,I}=1$, and $v_{\e \{j\},I}\not\equiv 0$.

\Bigdot $\ds S^I= \{\; h_l^I=
\hskip -.5cm \sum_{\tiny \begin{array}{c}
J\subset I,\,|J|\le l\\
\e_j=\pm 1, \, j \in J
\end{array}} \hskip -.25cm u_{J^c,l-|J|}\quad e^{-\b \t_{\e J}}\quad
v_{\e J,I},\,l\in \{1..|I|\}\; \}$ be a family of Poisson-commuting functions
($\ds \t_{\e J}=\sum_{j\in J}\e_j \t_j$).

If there exists a second set of  complex functions  $\tilde u_{K,p}$
obeying the first two conditions; 
such that $\ds \tilde S^I= \{\; \tilde h_l^I=
\hskip -.5cm \sum_{\tiny \begin{array}{c}
J\subset I,\,|J|\le l\\
\e_j=\pm 1, \, j \in J
\end{array}} \hskip -.25cm \tilde u_{J^c,l-|J|}\quad e^{-\b \t_{\e J}}\quad
v_{\e J,I},\,l\in \{1..|I|\}\; \}$ be a new family of Poisson-commuting functions;
and $\tilde u_{K,1}=u_{K,1}+c_1(|K|)$,
then there exist coefficients $c_r(m)$, $(r,m)\in \N^2$,
independent of all dynamical variables,
connecting the two families of Hamiltonians as:
$\tilde h_l^I=\sum_{s=0}^l c_{l-s}(|I|-s)\,h_s^I,
\,\mbox{ with }\, \forall m \in \N, c_0(m)=1$.
\end{theorem}

Hence the two relations for $p=0$ and $p=1$ between
the $U_{K,p}$'s and ${\cal U}_{K,p}$'s are sufficient to establish 
that the two sets of Hamiltonians define the same family of commuting 
dynamical 
flows, namely one set of Hamiltonians is a triangular linear combination
of the other set.
\section{Comments}

{\bf 1.} It follows from the uniqueness theorem that the \OSP RS hierarchy defined here
is equivalent to the KvD hierarchy for a particular set of couplings 
$\mu, \mu,\frac{\mu}{2}, \frac{\mu}{2}$. However as already emphasized the
full KvD pole structure is now obtained, contrary to the pure $BC_N$ case
where one coupling constant is actually set to zero. To obtain the complete KvD set of
hyperbolic potentials one clearly needs to define an ``extension'' (with the
same meaning as the ``extension'' of $BC_N$ CM models in~\cite{BSa,Ino} leading
to the full Inozemtsev potentials) of the \OSP Lax formulation. We hope to
report on this problem soon.
\vskip0.10in
\noindent {\bf 2.} In the elliptic case where $4$ coordinates $q$ are fixed to various combinations of the two
half-periods, the folding leads to a one-body potential with $4$ double poles located at
$q_i -$ (half-integer linear combinations of the two periods): indeed the term depending on $2q_i$
leads to a duplication of the $4$ poles induced by the $4$ fixed coordinates.

This is the correct denumbering and location of poles
in the classical limit of the quantum general elliptic Hamiltonian constructed in~\cite{Kom},
where $4$ pairs of poles separated by an order $\hbar$ degenerate to these $4$ double poles.
Of course, here again one does not obtain the full set of coupling constants (or
equivalently the residues at the poles). The conjecture
that folded RS models are the correct starting points for construction of a full
KvD-Hikami-Komori classical Lax-type representation may therefore be extended to
the elliptic case, with a suitably generalized underlying root lattice.
More precisely, we conjecture that this new root lattice is associated
with twisted toroidal superalgebras. Indeed it exhibits
three root lengths (pointing to superalgebras) with half-period
shifts (pointing to twisted algebras), precisely two independent
shifts of the shortest roots, interpreted as corresponding to the
two independent derivations characteristic of toroidal algebras.
\vskip0.10in
\noindent {\bf 3.} We have restricted ourselves here to the choice $\t_0 = \t_{\bar 0} = 0$.
In fact one can also choose either or both to be equal to a half-period $\frac {i \pi}{\beta}$.
The corresponding phase space manifold is also invariant under the Hamitonians $\H_l$. These
Hamiltonians therefore define an integrable hierarchy. At first sight it is not identical 
to the one defined here: the Hamiltonians indeed exhibit modifications of 
several relative signs in the potential terms; for instance the pure potential in the first Hamiltonian will 
become $\ds {\cal U}_{K,1}=
\prod_{\tiny \begin{array}{c}  k\in {\cal A}_{K}\\ k\not=0\end{array}}\hskip -.25cm v(q_k)
-  \prod_{\tiny \begin{array}{c}  k\in {\cal A}_{K}\\ k\not={\bar 0}\end{array}}\hskip -.25cm
 v(\frac{T}{2} -q_k)$ if $\t_{\bar 0} = \frac {i \pi}{\beta}$. Whether these
hierarchies are genuinely {\it new} integrable systems or may be obtained from our original
\OSP Hamiltonians by some redefinition of parameters is an open 
question which we will not address here.     

\section*{Acknowledgements}
We wish to thank Luc Frappat for his help in unraveling the meaning of the
underlying algebraic structure of the folding.


\end{spacing}
\end{document}